\documentclass[a4paper,12pt]{article}
\usepackage{graphicx}
\usepackage{setspace}
\usepackage{amsmath}
\usepackage[title]{appendix}
\usepackage[affil-it]{authblk}
\begin{document}

\title{Fireflies: New software for interactively exploring dynamical systems using GPU computing}
\author{Robert Merrison-Hort\thanks{Email: \texttt{robert.merrison@plymouth.ac.uk}}}
\affil{School of Computing, Electronics \& Mathematics, Plymouth University}
\date{}
%\email{robert.merrison@plymouth.ac.uk}
\maketitle

\begin{abstract}
In non-linear systems, where explicit analytic solutions usually can't be found, visualisation is a powerful approach which can give insights into the dynamical behaviour of models; it is also crucial for teaching this area of mathematics. In this paper we present new software, Fireflies, which exploits the power of graphical processing unit (GPU) computing to produce spectacular interactive visualisations of arbitrary systems of ordinary differential equations. In contrast to typical phase portraits, Fireflies draws the current position of trajectories (projected onto 2D or 3D space) as single points of light, which move as the system is simulated. Due to the massively parallel nature of GPU hardware, Fireflies is able to simulate millions of trajectories in parallel (even on standard desktop computer hardware), producing ``swarms'' of particles that move around the screen in real-time according to the equations of the system. Particles that move forwards in time reveal stable attractors (e.g. fixed points and limit cycles), while the option of integrating another group of trajectories backwards in time can reveal unstable objects (repellers). Fireflies allows the user to change the parameters of the system as it is running, in order to see the effect that they have on the dynamics and to observe bifurcations. We demonstrate the capabilities of the software with three examples: a two-dimensional ``mean field'' model of neuronal activity, the classical Lorenz system, and a 15-dimensional model of three interacting biologically realistic neurons.
\end{abstract}

\section{Introduction}
Many mathematical models are described by non-linear ordinary differential equations (ODEs). It is therefore important to be able to visualise solutions to these equations, and study how their dynamic behaviour changes when parameter values are modified. The theory of bifurcation analysis provides a rigorous mathematical framework for understanding how the stability of fixed points and limit cycles change as the parameters of the system are varied. Bifurcation theory can be combined with numerical continuation in order to trace the boundaries in parameter space that separate different dynamical regimes, and several software packages exist for this purpose, such as AUTO \cite{Doedel2012} and MATCONT \cite{Dhooge2003}. Before applying numerical continuation tools, in many cases it is useful to first get a rough understanding of how the phase space and system behaviour for particular parameter values, since the initial points for continuation (such as approximate fixed point and limit cycle positions) must be known. This is where more qualitative investigation (such as visualisation of the system) can be useful. 

In this paper we introduce new software, ``Fireflies''\footnote{Named after the classic dynamical systems example of firefly phase synchronisation, and also because the visualisations produced (slightly) resemble swarms of fireflies.} for the dynamic visualisation of ODE solutions. Instead of the traditional method of showing complete trajectories in phase space, Fireflies presents the user with a two- or three-dimensional view of a cloud of moving particles. Each particle represents the position in state space of one trajectory at the current point in time, and as the simulation runs the particles move around the screen according to the equations of the system. Different coloured groups of particles can be given different ranges of initial conditions, and each group can be either integrated forwards or backwards in time. Forwards moving particles illuminate attractors (e.g. stable fixed points and limit cycles) of the system, while backwards moving ones illuminate repelling (unstable) objects. By watching the motion of the particles, it is also possible to see features such as saddle separatrices, and how the speed of movement varies around limit cycles. Finally, Fireflies can also generate ``dynamic'' bifurcation diagrams, in which one of the visualised dimensions corresponds to a chosen parameter and each particle is given a different value for this parameter.

When only the current position of each trajectory is shown, a very large number of particles must be used in order for the structure of the system to be visible - we find that several million are required for good results. Numerically integrating this many equations quickly enough to display an interactive animation is not typically possible using the limited parallelism of traditional central processing units (CPU), but is an ideal problem for so-called ``massively parallel'' graphical processing units (GPUs). Unlike CPUs, which typically contain a small number (up to 18, at time of writing) of largely independent processing cores, GPUs consist of hundreds or thousands of cores, all of which execute the same code at the same time. Nowadays GPUs, which were originally developed to accelerate rendering in 3D graphics applications, sit alongside the central processing unit (CPU) in almost all modern computers, including tablets and smartphones. Despite their origins in the video games industry, the use of GPUs is rapidly becoming an essential part of many scientific computing applications, due to their low cost, widespread availability, and potential for large efficiency gains in certain types of computations. Fireflies consists of a user friendly graphical interface which can be used to produce simulations of N-dimensional systems of ODEs. Thanks to the power of GPU computing, these simulations can contain millions of particles while still running quickly enough to be interactive, allowing the user to change parameter values and immediately observe the effect of this on the particles' motion.

In this paper we will demonstrate the capabilities of Fireflies by showing visualisations of three example systems of ODEs. We will show how using Fireflies to study these systems can give insights into their behaviour and how their dynamics depends on their parameters. Note that the figures show static screenshots from moving simulations; in order to properly appreciate the capabilities of the software we recommend viewing the movies included in the Supplementary Material.

\section{Example Systems}
\subsection{A Two-Dimensional Model of the Basal Ganglia}
\label{sec:stngpe}
In this section we show an example of a two dimensional system of non-linear ODEs, corresponding to a simple model of neuronal activity. We show how Fireflies makes the dynamics of this system clear, and how the bifurcations of the system can be observed in an exciting new way. Although the system is two-dimensional, we also show how Fireflies can be used to create an interactive three-dimensional bifurcation diagram, by extending the visualisation into parameter space.

Based on earlier modelling work \cite{Gillies2002,Pavlides2012} we developed a mathematical model of activity in two connected regions in the brain: the subthalamic nucleus (STN) and external globus pallidus (GPe) \cite{Merrison-Hort2013}. These regions are both part of the ``basal ganglia'', a group of nuclei that are known to be involved in movement control and that are severely affected by Parkinson's Disease. A current subject of much interest is the fact that the Parkinsonian basal ganglia show a much larger degree of rhythmically modulated (oscillatory) activity \cite{Boraud2005}. We therefore use our model to study the conditions under which oscillations can appear in the STN-GPe network.

The equations of our model are:

\begin{equation}
\label{eqn:stn_gpe}
{\tau_s}\dot{x} = -x + {Z_s}(w_{ss}x - w_{gs}y + I)
\end{equation}
\begin{equation}
\label{eqnGPe}
{\tau_g}\dot{y} = -y + {Z_g}(-w_{gg}y + w_{sg}x)
\end{equation}

In these equations, which are based on the Wilson-Cowan formulation \cite{Wilson1972}, the variables $x$ and $y$ correspond to the average level of neuronal activity in the STN and GPe populations, respectively. The populations are coupled to each other and themselves through chemical connections (synapses), and the strength of these connections are given by the parameters $w_{gg}$, $w_{gs}$, $w_{ss}$ and $w_{sg}$ (so, for example, $w_{sg}$ is the strength of the connection from STN to GPe). The STN population is excitatory, meaning it acts to increase the activity of its synaptic targets, while the GPe is inhibitory, meaning it acts to decrease the activity of its synaptic targets. The functions $Z_s$ and $Z_g$ are monotonically increasing sigmoid curves that describe how each population responds to synaptic input, and the parameters $\tau_s$ and $\tau_g$ are time constants that determine how quickly the activity in each population changes.

Here we will show how Fireflies can be used to investigate how the behaviour of the system changes with one of its parameters. The parameter we will vary is $w_{ss}$, which corresponds to the strength of self-excitation in the STN. \cite{Gillies2002} demonstrated that if the neurons within the STN are able to excite each other (i.e. $w_{ss} > 0$) then the STN-GPe circuit is able to generate oscillations, although currently there is no known biological mechanism for such self-excitation. The simulations shown all contain two ``particle groups'', one of which is integrated forwards in time and other backwards. Green and pink particles are used to render the forwards and backwards particle groups, respectively. Currently Fireflies supports only one solver method, 4\textsuperscript{th} order Runge-Kutta, with a constant time step that can be adjusted by the user during the simulation. All particles are given random initial conditions, with values of the system variables drawn from a uniform random distribution on the interval $(0,1)$; the values 0 and 1 correspond to the minimum and maximum levels of population activity respectively. Any trajectories that leave this square region, or which have not been reset for more than time $T_{max}$, are reset to a new random set of initial conditions.

\begin{figure}[H]
\centering
\includegraphics[width=0.95\textwidth]{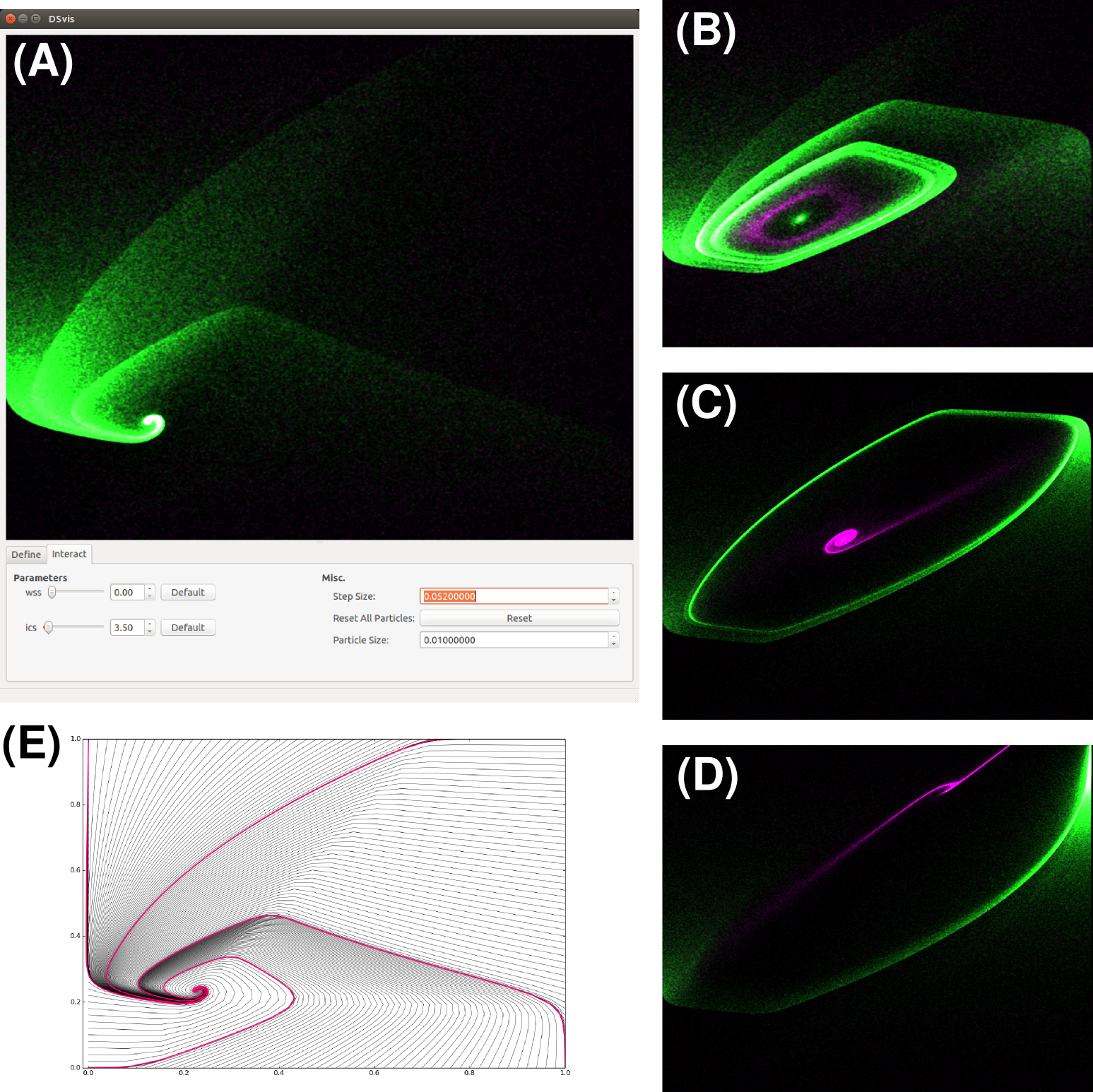}
\caption{Exploring the two dimensional STN/GPe model in Fireflies under variation of $w_{ss}$. (A)-(D) show screenshots from Fireflies for different values of $w_{ss}$, with 700,000 particles, half of which are integrated forwards in time (green) and half of which are integrated backwards (pink). (A) Full screenshot of the Fireflies window with $w_{ss}=0$. All forward-moving particles approach the globally stable spiral (in this still picture it is very difficult to see the backwards particles, which diverge quickly to infinity). (B) $w_{ss}=4.9$: A stable spiral, unstable limit cycle, and stable limit cycle. (C) $w_{ss}=7.8$: An unstable spiral and a stable limit cycle. (D) $w_{ss}=11$: An unstable node, saddle (not visible), and stable node. (E) Phase portrait (not generated by Fireflies) of the STN/GPe system with $w_{ss}=0$. All trajectories start from the borders of the phase plane and spiral in to the fixed point. The trajectories that start in the four corners of the phase plane are shown in pink.}\label{fig:stngpe_composite}
\end{figure}

Figure \ref{fig:stngpe_composite}A shows a screenshot of the simulation when $w_{ss}=0$, corresponding to the situation with no STN self-excitation. Under these conditions there is a single stable fixed point, which all of the green (forwards time) particles spiral in towards. The pink (backwards time) particles are difficult to see in this image, as they move very quickly out of the bounds of the system (to infinity) and are constantly being reset to new random initial conditions. It is interesting to note from this image that the density of particles approaching the spiral is not uniform, and has several discontinuities where the density suddenly changes. We suspected that this phenomena is a result of the boundaries of the phase space, and confirmed this with a traditional phase portrait where all the trajectories began on the edges of phase space (Figure \ref{fig:stngpe_composite}B). The borders where particle density changes quickly correspond to the trajectories that begin in the four corners of phase space (red lines in Figure \ref{fig:stngpe_composite}B); these four trajectories divide the phase plane into regions that ``funnel'' trajectories into the spiral.

The slider controls (visible in the bottom left of Figure \ref{fig:stngpe_composite}A) can be used to explore the effects of changing the parameters of the system. When the value of a parameter is changed using its slider, the dynamics of the system change immediately and the difference can be clearly seen in the movement of the particles. Figure \ref{fig:stngpe_composite}C shows the STN/GPe system after the strength of STN self-excitation ($w_{ss}$) has been increased to 4.9, taking the system through a fold of limit cycles bifurcation. For this parameter value, a pair of limit cycles (stable and unstable) now encircle the original stable fixed point. This illustrates the purpose of the group of particles that are integrated backwards (pink): these particles are attracted to unstable objects, here they reveal the location of the unstable limit cycle. As $w_{ss}$ is increased further, the unstable limit cycle shrinks around the fixed point and eventually disappears in a subcritical Andronov-Hopf bifurcation, at which point the fixed point becomes unstable. After this bifurcation the limit cycle is globally stable, but as $w_{ss}$ is increased further the period of oscillation increases. This can be seen on the visualisation in the form of particles ``bunching up'' and moving very slowly around one part of the cycle (Figure \ref{fig:stngpe_composite}C). Eventually, a saddle-node on invariant circle (SNIC) bifurcation occurs, after which all trajectories approach a new, globally stable, fixed point.

\begin{figure}[H]
\centering
\includegraphics[width=0.95\textwidth]{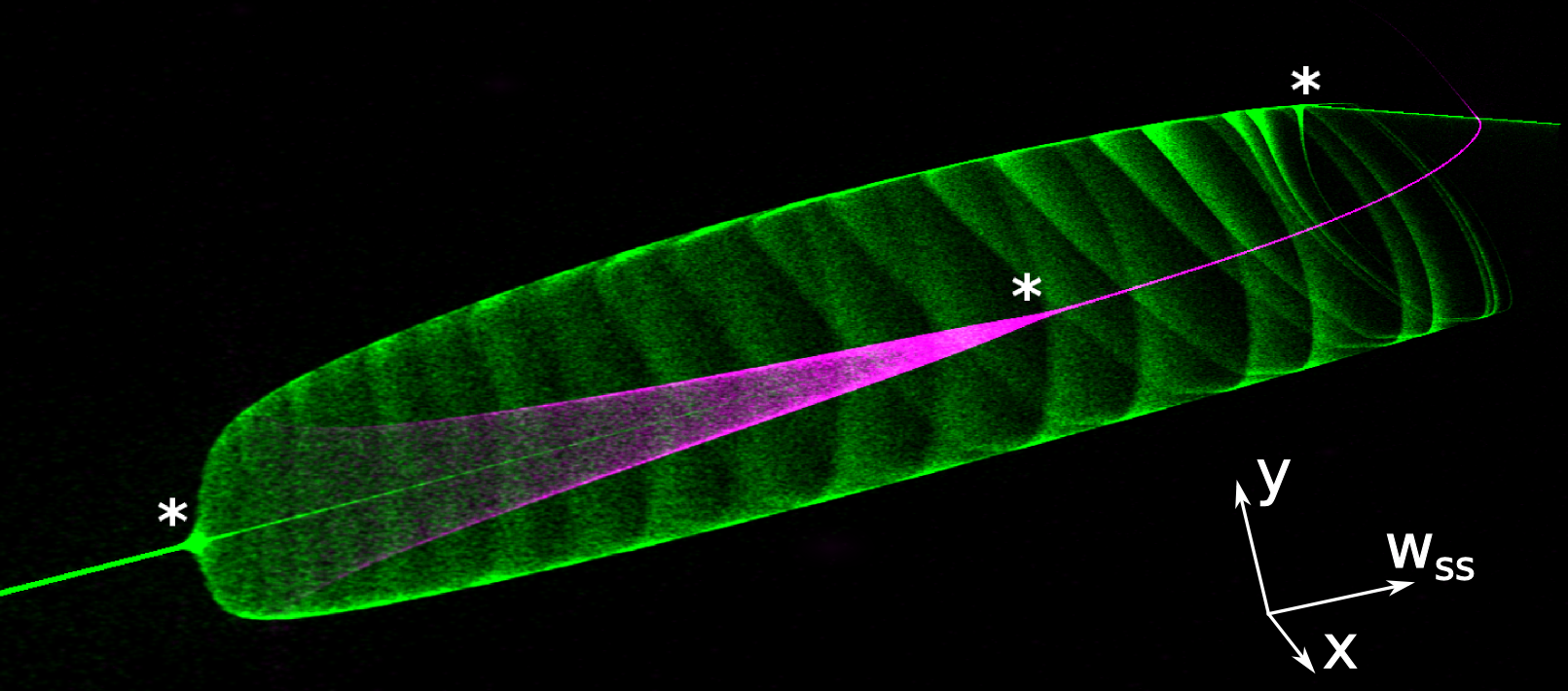}
\caption{3D bifurcation diagram of the STN/GPe system under variation of the parameter $w_{ss}$. The simulation contains 8 million particles, divided into a forwards in time group (green) and a backwards in time group (pink). The three visible bifurcations are marked with asterisks, from left to right: fold of limit cycles, subcritical Andronov-Hopf, saddle node on invariant circle. Note how the particles on the stable limit cycle become increasingly ``bunched'' at the top as the cycle approaches the SNIC bifurcation. The asterisks and direction arrows were added to the image manually and were not generated by Fireflies.}\label{fig:stngpe_3d}
\end{figure}

Fireflies can also be used to generate three-dimensional animated bifurcation diagrams of two-dimensional systems. To see this, we redefine the system so that the bifurcation parameter ($w_{ss}$) is a new state variable, subject to $dw_{ss}/dt=0$. The initial condition range for this new state variable is set to be the range of parameter values that are of interest. With this configuration, each particle has its own value for $w_{ss}$, and the space consists of a ``continuum'' of phase portraits, beautifully illustrating the dynamics of the system (Figure \ref{fig:stngpe_3d}). The user can further investigate the effects of the other parameters on the system's dynamics by varying them interactively and observing how this changes the 3D bifurcation diagram.

\subsection{The Lorenz Equations}
In this example we use Fireflies to visualise the dynamics of the three-dimensional Lorenz system. We describe in detail the results of various visualisations obtained for different values of a parameter of the system, and then produce an animated bifurcation diagram that summarises these results.

The classical Lorenz system consists of three coupled non-linear differential equations:
\begin{equation}
\label{eqn:lorenz_x}
\dot{x} = \sigma (y-x)
\end{equation}
\begin{equation}
\label{eqn:lorenz_y}
\dot{y} = x(r - z) - y
\end{equation}
\begin{equation}
\label{eqn:lorenz_z}
\dot{z} = xy - \beta z
\end{equation}

Where $\sigma$, $r$ and $\beta$ are parameters of the system ($\sigma,r,\beta > 0$). These equations were originally studied as a simplified model of convection in the atmosphere \cite{Lorenz1963}, and a straightforward description of the different dynamical regimes that they can produce, including chaos, can be found in any textbook on nonlinear dynamics (e.g. \cite[pp.311-320]{Strogatz1994}). In this section we will demonstrate the ability of Fireflies to visualise three dimensional systems by exploring the behaviour of the Lorenz equations in the case where $\sigma=10$ and $\beta=\frac{8}{3}$, while the parameter $r$ is gradually increased from zero.

\begin{figure}[H]
\centering
\includegraphics[width=0.95\textwidth]{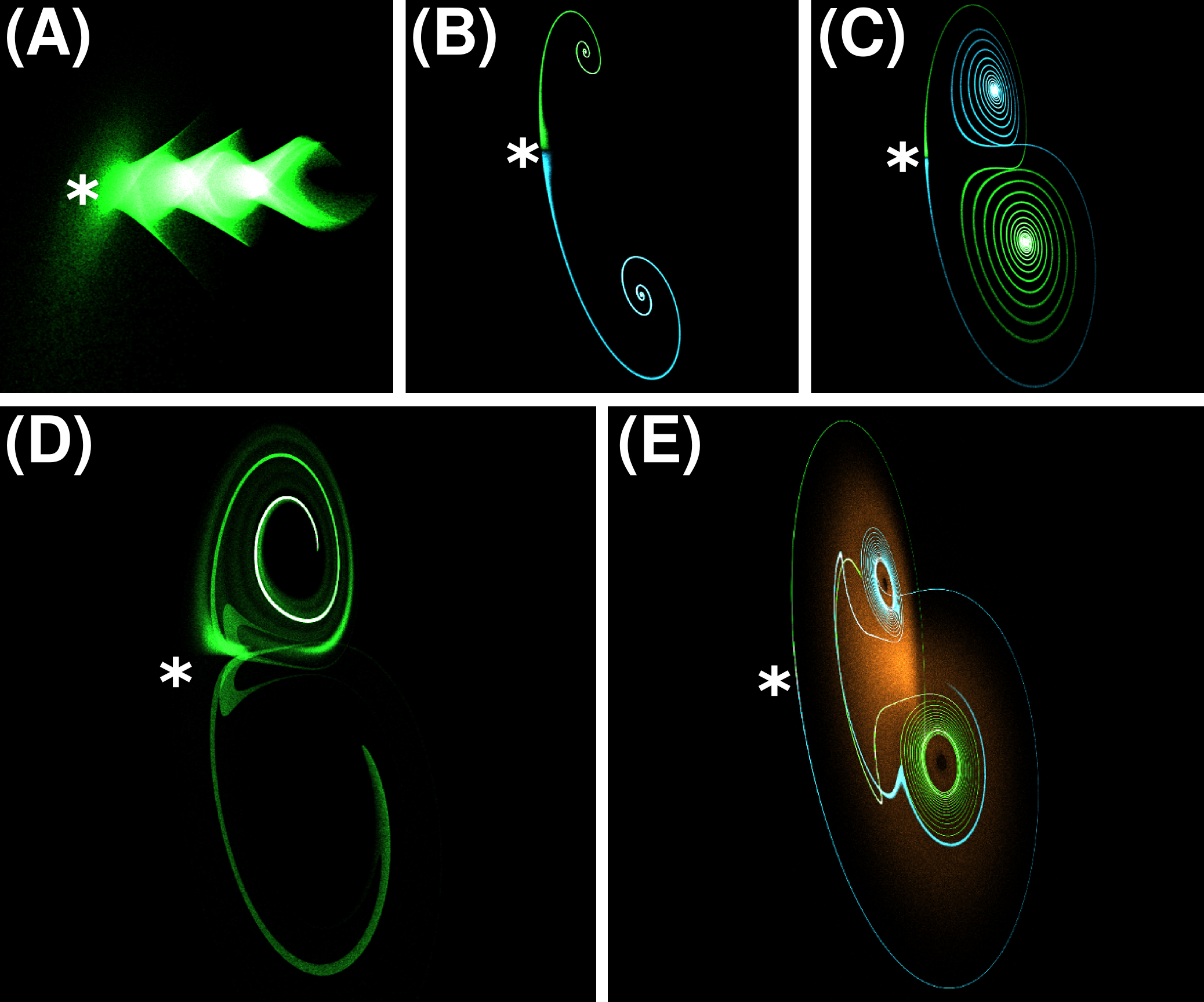}
\caption{Exploring the Lorenz equations by varying $r$. All particles are integrated forwards in time, different colours use different initial condition ranges. Asterisks approximately indicate the origin. (A) $r=0.5$: All particles approach the stable fixed point at the origin. The complex shape is due to the boundaries of the initial conditions ($x\in(-10,10), y\in(-30,30), z\in(0,50)$) and the shape of the fixed point's manifolds. (B) $r=5$: The origin is now a saddle and two new stable fixed points have emerged. Both groups of particles have initial conditions near to the origin, but on opposite sides of its incoming separatrix (green: $x,y\in(0,0.01), z\in(0,0.01)$, blue: $x,y\in(-0.01,0), z\in(0,0.01)$). Trajectories approach one of the spirals, based on which side of the separatrix they start on. (C-D) $r=15$: Trajectories now switch between the two spirals for some time, before settling down and approaching one of them (transient chaos). (C) has initial conditions near the saddle, as in (B); these trajectories loop once around one spiral before approaching the other. (D) has initial conditions further away from the saddle; these trajectories can switch sides repeatedly. (E) $r=25$: A strange attractor. Trajectories starting near the saddle (green and blue) begin by approaching one of the spirals, before slowly spiralling away and swapping sides chaotically. Another group of particles (bronze) with initial conditions spanning a large area show the shape of the attractor.}\label{fig:lorenz_composite}
\end{figure}

For values of $r$ less than 1, the origin is the only stable fixed point. Figure \ref{fig:lorenz_composite}A shows Fireflies when $r = 0.5$, shortly after the simulation has started and all the particles are quickly approaching the origin. As was also seen in the 2D system described above, Fireflies shows how the space occupied by the cube of initial conditions is deformed around the fixed point as the particles move. Note that we do not show any backwards-moving particles in this section, as they are not generally useful in visualisations of the Lorenz equations. This is because any unstable fixed points or cycles are of saddle type, which means that neither forward nor backward moving particles tend to them as $t\rightarrow\infty$. 

At $r=1$ a supercritical pitchfork bifurcation occurs, and two new stable fixed points emerge from the origin. Figure \ref{fig:lorenz_composite}B shows a screenshot from Fireflies with $r = 5$. With the new parameter value the origin becomes a saddle, and particles begin to move away from it in one of two directions, spiralling in to one of the new stable fixed points that were created in the bifurcation. To clearly show how trajectories spiral in to the fixed point, this simulation has two groups of particles (green and blue) with initial conditions that are drawn from separate small volumes in state space, both very close to the saddle at the origin, but on opposite sides of its incoming separatrix. The paths that the particles in these two groups take away from the saddle are very close to the saddle's outgoing separatrices. Trajectories starting at other points in state space all approach the spiral that is on the same side of the saddle's incoming separatrix as their initial position.

At $r \approx 13.926$ a homoclinic bifurcation occurs, at which point the saddle's outgoing separatrices join up with its incoming ones, resulting in the creation of a pair of unstable limit cycles. For values of $r$ beyond the bifurcation the separatrices have ``crossed over'', and trajectories that begin near the saddle make one cycle around their nearest spiral before returning back towards the saddle, and then spiralling in to the stable fixed point on the opposite side of the incoming separatrix, as shown in Figure \ref{fig:lorenz_composite}C. Trajectories that start a bit further away from the saddle, however, rotate around their closest spiral at a much greater distance. When this rotation brings them close to the saddle's incoming separatrix, some of them split off and begin rotating around the opposite spiral. This unpredictable swapping, which corresponds to transient chaos, can happen many times before a particle finally spirals fully in to one of the two stable fixed points. Figure \ref{fig:lorenz_composite}D shows one group of particles which all start closer to the top spiral than the bottom one, but a little further away from the origin than those in Figure \ref{fig:lorenz_composite}C. Although most spiral in to the top fixed point, with each rotation a mass of particles splits off and switches to rotate around the bottom spiral.

Finally, at $r\approx24.06$ a strange attractor appears. Now, even though the two spirals remain stable until $r\approx24.74$, almost all trajectories flip backwards and forwards between the two spirals infinitely and chaotically. In Figure \ref{fig:lorenz_composite}E, the two groups of particles that start near to the origin (green and blue) initially behave as in \ref{fig:lorenz_composite}C, performing one large loop around their closest spiral before approaching quite close to the opposite spiral. The spirals are only very weakly stable now, however, and the trajectories spiral slowly away from them, instead of into them. After some time (as is just beginning to happen in \ref{fig:lorenz_composite}E), the particles come far enough away from the spiral that some of them switch sides, beginning an endless series of such seemly random side swappings on the strange attractor. The figure also contains a third set of particles (bronze coloured) which start from a much wider set of initial conditions; these show the general shape of the strange attractor's surface.

\begin{figure}[H]
\centering
\includegraphics[width=0.95\textwidth]{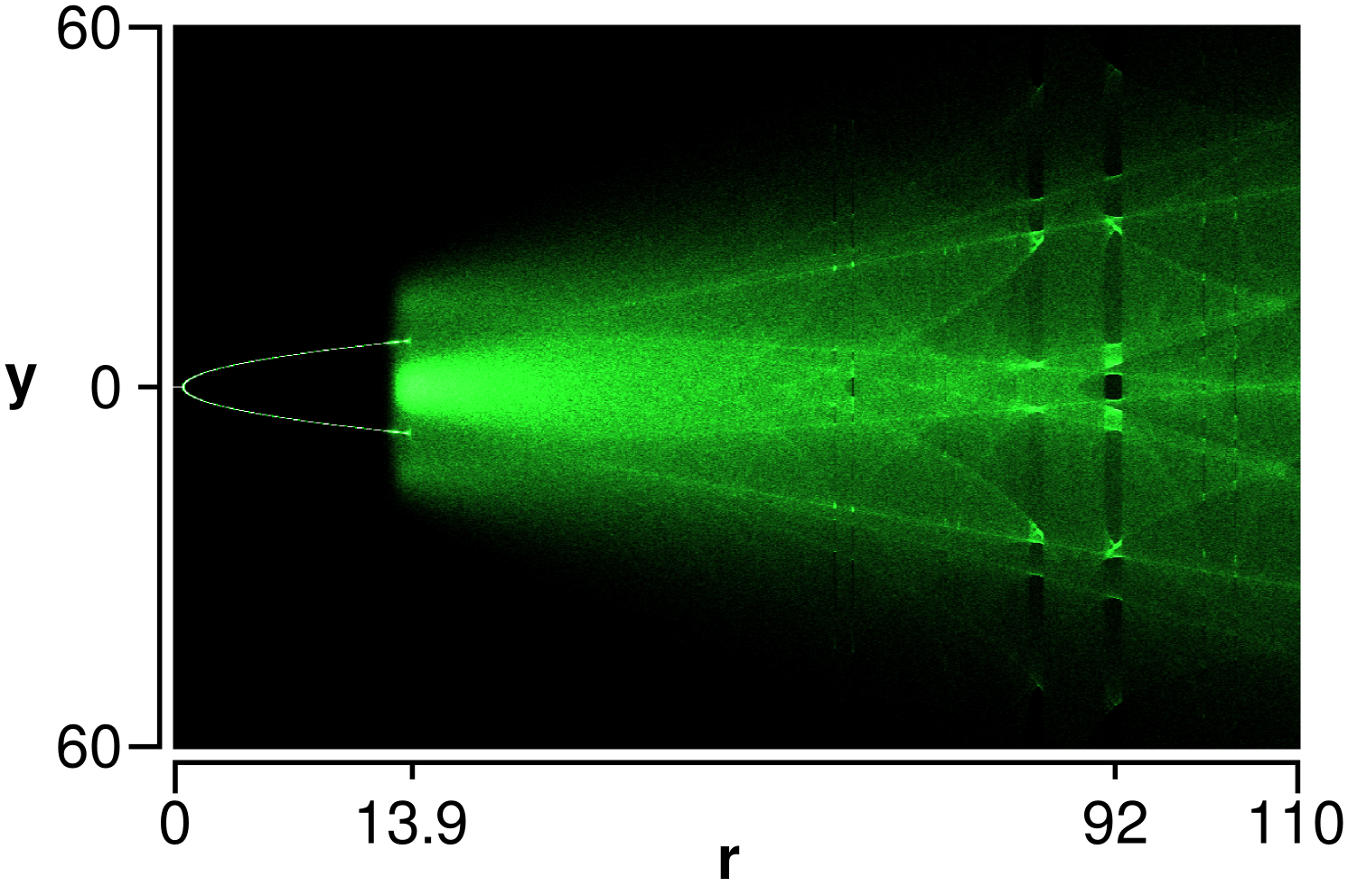}
\caption{A ``live'' bifurcation diagram for the Lorenz equations. The diagram was created by setting up a simulation with 8 million particles, each with a random fixed value for $r$, projected onto the $(r, y)$ plane. The supercritical pitchfork at $r=1$ is clearly visible, as is the appearance of a strange attractor at $r\approx13.9$ and various transient windows of non-chaotic behaviour such as at $r=92$. Axes added manually.}\label{fig:lorenz_bif}
\end{figure}

By applying the same technique as in the previous section, we can also use Fireflies to generate an animated bifurcation diagram for the Lorenz equations. To achieve this, we make parameter $r$ a state variable with $\dot{r}=0$ and set up a two dimensional projection with axes $r, y$. Note that a three dimensional projection could also be used, but due to the perspective transformation the results are not shown as clearly in this case. Figure \ref{fig:lorenz_bif} shows a screenshot from a simulation using this configuration, where each particle's initial value of $r$ is chosen from the interval $(0\cdots110)$. The supercritical pitchfork bifurcation is clearly visible at $r=1$, and the appearance of a strange attractor at $r\approx13.9$ is marked by particles beginning to form a cloud that resembles noise. This cloud clearly has some very detailed structure, however, as can be seen more and more clearly as $r$ increases. Several small windows of parameter values where the dynamics become regular are also visible, for example at $r=92$. Again, it should be kept in mind that Figure \ref{fig:lorenz_bif} is a static screenshot of a moving animation of 8 million particles. The other two parameters of the system can be changed interactively, and the effect of this variation on the bifurcation diagram is seen immediately.

\section{Coupled Hodgkin-Huxley Neurons}
\label{sec:hh}
In this section we demonstrate the use of Fireflies to visualise a system that is considerably more complex than those presented above, consisting of 15 coupled ODEs that describe the activity in three synaptically connected neurons. 

In 1963 Alan Hodgkin and Andrew Huxley received a Nobel prize for their discovery of the mechanism that allows neurons to generate action potentials (the electrical impulses, or ``spikes'', that neurons use to communicate with each other). More than 50 years later the general structure of the equations that Hodgkin and Huxley used to describe this mechanism still forms the basis of an enormous number of biologically realistic computational models of neuronal activity. Here, however, we will use the original equations and parameters, which specifically relate to biophysical activity in the giant axon of the squid \cite{Hodgkin1952}. In this model, the electrical potential across a part of the neuron's membrane evolves according to the flow of sodium (Na) and potassium (K) ions through the membrane. The permeability of the membrane to these ions is controlled by gates, which open and close according to the membrane potential. Equations \ref{eqn:hh_v}--\ref{eqn:hh_n} describe the ``classical'' Hodgkin-Huxley model of neuronal dynamics in a population of $N$ neurons.

\begin{equation}
\label{eqn:hh_v}
C\dot{V_i} = \bar{g}_{lk}(e_{lk}-V_i) + h_i{m_i}^3\bar{g}_{na}(e_{na}-V_i) + {n_i}^4\bar{g}_k(e_k-V_i) + {I^i}_{syn}(t) + I_i
\end{equation}
\begin{equation}
\label{eqn:hh_h}
\dot{h_i} = \alpha_h(V_i)(1-h_i) - \beta_h(V_i)h_i
\end{equation}
\begin{equation}
\label{eqn:hh_m}
\dot{m_i} = \alpha_m(V_i)(1-m_i) - \beta_m(V_i)m_i
\end{equation}
\begin{equation}
\label{eqn:hh_n}
\dot{n_i} = \alpha_n(V_i)(1-n_i) - \beta_n(V_i)n_i
\end{equation}
\begin{equation*}
i = 1, 2, ..., N
\end{equation*}

Here $V_i$ is the membrane potential of the $i^{th}$ neuron and $h_i$, $m_i$ and $n_i$ represent the average state of the gates on its ion channels ($0\leq h_i,m_i,n_i \leq1$). The parameters $C$, $\bar{g}_{lk}$, $\bar{g}_{na}$, $\bar{g}_k$, $e_{lk}$, $e_{na}$ and $e_k$, along with the functions $\alpha_X(.)$ and $\beta_X(.)$ ($X \in \{h,m,n\}$), are described in \cite{Hodgkin1952} or any computational neuroscience textbook. All parameters mentioned so far take the values given in \cite{Hodgkin1952}. Parameter $I_i$ controls how much external current is injected into the cell - increasing this parameter causes a transition from quiescence to single action potential firing to repetitive firing. Finally, ${I^i}_{syn}$ represents the total synaptic current that arises as a result of inputs from the other neurons in the model. If $I_i={I^i}_{syn}=0$ then the membrane potential of neuron $i$ approaches a fixed point at the ``resting'' potential of 0 (mV). In this example we use a network where the neurons are arranged in a loop, with each receiving synaptic input from only the previous neuron:

\begin{equation}
\label{eqn:hh_isyn}
{I^i}_{syn}(t)=\bar{g}_{syn}(e_{syn}-V_i)s_{(i-1) mod N}
\end{equation}
\begin{equation}
\label{eqn:hh_s}
\dot{s_i}=\tau_r^{-1}(1+exp(-\sigma(V_i-\theta)))^{-1}(1 - s_i) - \tau_d^{-1}s_i
\end{equation}

The parameter $\bar{g}_{syn}$ is the maximum conductance (mS) of a synaptic connection (its ``strength''); each synaptic connection has the same strength in this model. $e_{syn}$ is the synaptic equilibrium potential: if $e_{syn}>0 (mV)$ then synaptic input is excitatory and acts to raise the membrane potential of the post-synaptic neuron above the resting state; in this example we set $e_{syn}=10$. For each neuron we consider a variable $s_i$, where $0<s_i<1$, which acts as an indicator for when the neuron is spiking. This variable increases to 1 very rapidly (with time constant $\tau_r$) when the neuron is firing an action potential, and then decays slowly (time constant $\tau_d$) after a spike, as described by equation \ref{eqn:hh_s}. Parameters $\sigma$ and $\theta$ are the slope and shift respectively of the sigmoid function which is used to increase $s_i$.

\begin{figure}[H]
\centering
\includegraphics[width=0.95\textwidth]{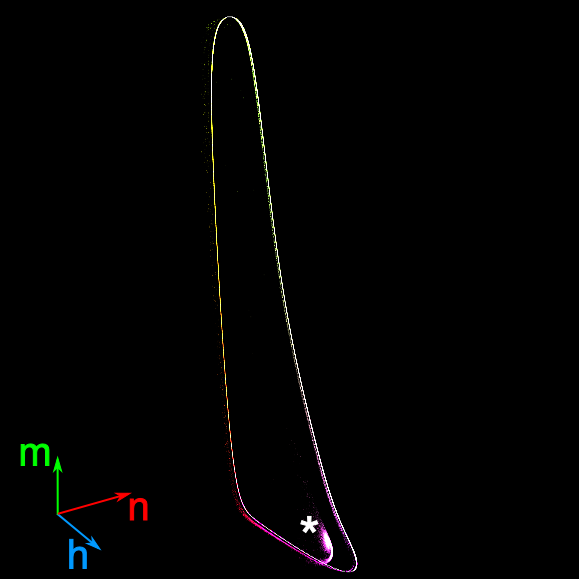}
\caption{A screenshot of Fireflies running the single Hodgkin-Huxley neuron model with 500,000 particles. The position of each particle is projected into the 3D space $(h_1,m_1,n_1)$. Particles are coloured according to their position on each axis. Asterisk indicates the approximate position of the (weakly) unstable fixed point, with a cloud of particles moving slowly away from it. Axes added manually.}\label{fig:hh_single}
\end{figure}

We begin by briefly showing a visualisation of the four dimensional phase space of a single independent neuron under varying strengths of current injection. For low values of $I_1$ all trajectories approach the fixed point which corresponds to the resting state. As $I_1$ is increased past $I_1 \approx 6.25$ the resting state undergoes an Andronov-Hopf bifurcation and a stable limit cycle appears. This limit cycle corresponds to periodic spiking with a frequency that increases with $I_1$. Figure \ref{fig:hh_single} shows a screenshot of a simulation with a single neuron and $I_1=10$. The three dimensions of the projection are the three gating variables: $h_1$, $m_1$ and $n_1$. All the particles in this simulation eventually approach the stable limit cycle. However, there is also a spiral-shaped cloud of particles (marked with an asterisk), which is made up of trajectories that start close to the formerly stable fixed point. For parameter value $I_1=10$ this fixed point is only weakly unstable, so trajectories nearby spend a long time oscillating with a low (gradually increasing) amplitude before they approach the limit cycle.

\begin{figure}[H]
\centering
\includegraphics[width=0.95\textwidth]{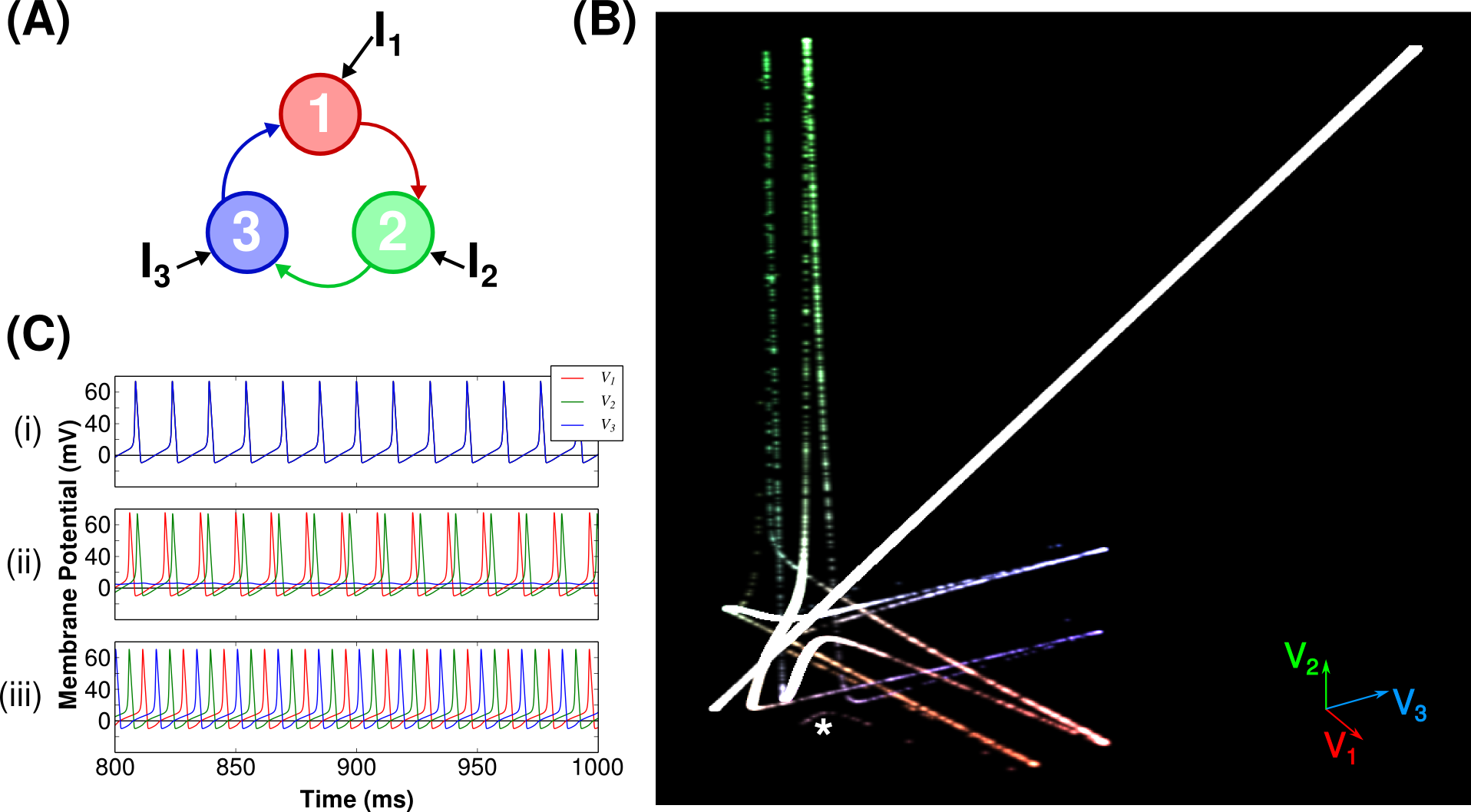}
\caption{A Hodgkin-Huxley model with three neurons spiking repetitively in response to current injection, connected by weak excitatory synapses. (A) An overview of the model's structure. (B) A screenshot of Fireflies running the model with 500,000 particles. The position of each particle is projected into the 3D space $(V_1,V_2,V_3)$. Particles are coloured according to their position on each axis. Asterisk indicates a small group of particles that are close to the limit cycle where each neuron spikes in turn. Axes added manually. (C) Individual plots (from XPPAUT) of trajectories starting from three different sets of initial conditions, leading to different limit cycles. i: Synchronous spiking; ii: Two neurons spiking, one suppressed; iii: Neurons spiking in turn.}\label{fig:hh_3}
\end{figure}

Next, we consider a ring of $N=3$ neurons (Figure \ref{fig:hh_3}A), where each neuron receives an identical current injection that is strong enough to cause repetitive firing ($I_{j}=10,j=1,2,3$). Such networks of coupled regular spiking neurons are typically able to synchronise their firing in either in-phase or anti-phase, depending on the strength and nature of synaptic connections \cite{Mirollo1990}. We set $e_{syn}=10 (mV)$ and $\bar{g}_{syn}=0.5 (mS)$ to simulate the case of weak excitatory coupling. From running a few simulations of the system from random initial conditions in XPPAUT \cite{Ermentrout2002} it was clear that with these parameter values there was a very stable synchronous state, where all three neurons fired regularly with the same frequency and phase shift. Figure \ref{fig:hh_3}B shows a screenshot of Fireflies after this system has been simulated for long enough that all particles have settled down onto stable attractors. Each particle's position projected onto the space $(V_1,V_2,V_3)$. The colour of each particle varies with its position,using a projection from the three dimensional co-ordinates to an RGB colour code. The prominenet solid white diagonal line is the synchronous limit cycle; most particles are attracted to this and continually move along the diagonal as the three neurons spike in unison. Figure \ref{fig:hh_3}C(i) shows the synchronous state as produced by XPPAUT.

In addition to the synchronous state, however, the visualisation also reveals four other stable limit cycles, three of which are easily visible with 500,000 particles. The three limit cycles appear as six coloured prongs in Figure \ref{fig:hh_3}B. By interactively moving around 3D space we were able to see that the particles were organised into three closed orbits, each of which included two differently coloured prongs. Each of these closed orbits corresponds to a limit cycle where two of the neurons are spiking and the third is silent. Simulations in XPPAUT revealed that each of these limit cycles corresponds to the state where the pre-synaptic neuron fires first, followed by the post-synaptic neuron, followed by a pause before the cycle repeats; this is shown in Figure \ref{fig:hh_3}C(ii). This detail about the order in which the neurons fire is not immediately visible in the Fireflies visualisation.

The other stable limit cycle that we have found has a relatively small basin of attraction, meaning that not many of the 500,000 particles approach it. It is also quite difficult to see in still screenshots, although some particles on this cycle are marked with an asterisk in Figure \ref{fig:hh_3}B. By following the path of these particles around the visualisation space, we can see that their trajectories correspond approach a stable limit cycle corresponding to the state where all three neurons spike in sequence. Interestingly, this order of this sequence is opposite to that of the direction of synaptic coupling (i.e. the firing sequence is neuron 1, neuron 3, neuron 2, ...). In order to verify that this was indeed a real limit cycle and not a numerical error in Fireflies, we wrote a short script to repeatedly (and sequentially) run simulations from random initial conditions in XPPAUT. The script recorded any sets of initial conditions that lead to solutions where all the neurons fired repetitively but non-synchronously. After performing a very large number of simulations this script had found only two sets of initial conditions that lead to the limit cycle that we had identified in Fireflies, and the resulting plots of $V(t)$ are shown in Figure \ref{fig:hh_3}C(iii).

The example in this section demonstrates a case where Fireflies could reveal several stable dynamical regimes that were not immediately obvious from running individual simulations of the system from random initial conditions. Due to the fact that the system under study is composed of three identical elements, it is not surprising to find that multiple regimes corresponding to different symmetries are possible. Furthermore, it is likely that other limit cycles exist - for example there may be one in which all three neurons fire in sequence in order of synaptic coupling, rather than in the opposite order as we observed here. However, because none of the hundreds of thousands of particles in our simulations approached such attractors, it is likely that they are either unstable or have extremely small basins of attraction. 

\section{Discussion}
To facilitate the qualitative study of systems of ODEs, software such as XPPAUT \cite{Ermentrout2002}, often includes the ability to plot trajectories in phase space that start from multiple sets of initial conditions, chosen either from a regular grid in phase space or stochastically. However, in systems with high dimensional phase space (many ODEs) or with finely structured dynamical regimes, a very large number of initial conditions must be used in order to have a high level of confidence that all stable attractors have been found. This can make the qualitative investigation a very slow process, especially if one wishes to also examine how the picture varies with parameter values. It is also very difficult to visualise many different trajectories on the same phase portrait, as large numbers of them will completely fill the phase space. We believe that Fireflies offers a useful and exciting new way to investigate dynamical systems. This approach is more intuitive than traditional phase portraits, as it presents dynamical systems as they are: \textit{dynamic}.

In this paper we have demonstrated, through several examples, the power of Fireflies for qualitatively exploring the behaviour of dynamical systems. The visualisation of systems in explorable 2D or 3D space provide a new perspective on such systems that can help with intuitively understanding them. In addition to this, since Fireflies can simulate millions of trajectories in parallel, stable attractors can be discovered that might not have been found with other methods of qualitative investigation - as in section \ref{sec:hh} (although clearly as the dimensionality of systems increases, the number of particles required to uniformly sample phase space with a given resolution increases exponentially). 

\begin{table}
\centering
\begin{tabular}{c|c|cc|c}
   & & \multicolumn{2}{c|}{CPU} & \\ 
  System & Particles & Single Core& Quad Core (est.) & GPU \\ \hline \hline
  STN-GPe & 700k &182ms & 46ms & 1ms \\
  Lorenz & 3m & 109ms & 27ms & 3ms \\
  Hodgkin-Huxley & 500k & 942ms & 236ms & 22ms \\ 
\end{tabular}
\caption{The time taken to advance each of the systems shown in this paper by a single step (averaged over 1000 steps), when executed on a single core CPU and a GPU. The estimated best-case performance for a quad-core CPU was estimated by dividing the single core figure by four. The CPU used was an Intel Core i5-2500k (3.3Ghz) and the GPU was an Nvidia Geforce GTX 460. The C code for the CPU benchmark (available at www.bitbucket.org/rmerrison/fireflies) was adapted from the corresponding OpenCL kernel for each system, and was compiled using GCC 4.8.2 on Ubuntu 14.04 using the \texttt{-Ofast} option.}
\label{tab:benchmark}
\end{table} 

Scientific computing is increasingly taking advantage of the power and availability of GPUs, and a range of techniques have emerged for using GPUs to solve problems in domains such as computational neuroscience \cite{Brette2012} and fluid dynamics \cite{Harris2004}. To demonstrate the necessity of ``massively parallel'' GPU computing for producing Fireflies' visualisations, we compared the average time taken to compute a single integration step using a traditional CPU (single core and estimated quad core) to the time taken when integration is performed on the GPU. Table \ref{tab:benchmark} shows the results of this comparison. For interactive graphical applications, it is normally considered necessary to render at least 30 frames per second in order to give the user a smooth experience, which means that each frame must be generated in 33ms or less. Since the times shown in table \ref{tab:benchmark} only represent the time taken to advance the simulation and do not include additional time needed to render particles to the screen, it would seem to be very difficult or impossible to run visualisations shown in this paper at an acceptable frame rate without using the GPU. Additionally, running the simulation on the CPU would also incur the significant additional overhead of passing the particles' current positions to the computer's graphical hardware for rendering; by performing integration on the GPU Fireflies minimizes the number of CPU-GPU memory transfers.

We have found that our new visualisation technique can be particularly useful in a teaching context, as it very easily demonstrates the behaviour that a particular set of equations produces. For example, visualisation of simple systems, such as the two-dimensional model of neuronal activity illustrated in section \ref{sec:stngpe}, can be used to show different bifurcations in a very intuitive way. For example, the parameter $w_{ss}$ is increased, Fireflies shows how particles move more and more slowly around one part of the limit cycle, becoming increasingly bunched together until finally being attracted into a new stable fixed point that appears in a SNIC bifurcation.

The ability of Fireflies to quickly simulate millions of trajectories at the same time is due to the fact that almost all processing occurs exclusively within the GPU, but this approach carries a number of disadvantages. In particular, due to the time taken to transfer data between GPU and CPU it is not possible to permanently record trajectories (e.g. for further analysis), however in future we plan to add the ability to record either a subset of trajectories, or all trajectories but with a recording interval that is greater than the time step. It should be noted, however, that this is not the main intended purpose of Fireflies, and other libraries for exploiting GPUs to solve ODEs, such as the Odeint library for C++ \cite{Ahnert2011}, may be more appropriate. Another limitation that arises from the use of the GPU is that Fireflies does not allow the equations of the system to contain branching (e.g. changing variable values in response to threshold crossing). This is because GPUs are based on a ``single instruction multiple data'' (SIMD) architecture, whereby each processing core executes the same instruction at the same time; due to this architecture branching causes a significant reduction in computation speed.

Fireflies is written in Python and utilises a number of cross-platform and open source libraries, namely NumPy \cite{VanDerWalt2011}, PyOpenGL\footnote{http://pyopengl.sourceforge.net/}, PyOpenCL \cite{Klockner2012}, PySide\footnote{http://www.pyside.org} and Mako\footnote{http://www.makotemplates.org}. The software has been tested in Linux and Windows, and should also work in Apple OSX. The full source code for Fireflies, along with instructions for its use, can be found at https://bitbucket.org/rmerrison/fireflies.

\section*{Acknowledgement}
I am extremely grateful to Roman Borisyuk for the valuable input that he gave during the development of Fireflies and the preparation of this manuscript.

\bibliographystyle{ieeetr}
\bibliography{main}

\begin{thebibliography}{10}

\bibitem{Doedel2012}
E.~J. Doedel and B.~E. Oldeman, ``Auto-07p: Continuation and bifurcation
  software for ordinary differential equations,'' tech. rep., Concordia
  University, Montreal, 2012.

\bibitem{Dhooge2003}
A.~Dhooge, W.~Govaerts, and Y.~A. Kuznetsov, ``Matcont: A matlab package for
  numerical bifurcation analysis of odes,'' {\em ACM Transactions on
  Mathematical Software}, vol.~29, no.~2, pp.~141--164, 2003.

\bibitem{Gillies2002}
A.~Gillies, D.~Willshaw, and Z.~Li, ``Subthalamic-pallidal interactions are
  critical in determining normal and abnormal functioning of the basal
  ganglia,'' {\em Proceedings of the Royal Society of London. Series B:
  Biological Sciences}, vol.~269, no.~1491, pp.~545--551, 2002.

\bibitem{Pavlides2012}
A.~Pavlides, S.~John~Hogan, and R.~Bogacz, ``Improved conditions for the
  generation of beta oscillations in the subthalamic nucleus-globus pallidus
  network,'' {\em European Journal of Neuroscience}, vol.~36, no.~2,
  pp.~2229--2239, 2012.

\bibitem{Merrison-Hort2013}
R.~J. Merrison-Hort, N.~Yousif, F.~Njap, U.~G. Hofmann, O.~Burylko, and
  R.~Borisyuk, ``An interactive channel model of the basal ganglia: Bifurcation
  analysis under healthy and parkinsonian conditions,'' {\em The Journal of
  Mathematical Neuroscience}, vol.~3, no.~1, 2013.

\bibitem{Boraud2005}
T.~Boraud, P.~Brown, J.~Goldberg, A.~Graybiel, and P.~Magill, ``Oscillations in
  the {B}asal {G}anglia: The good, the bad, and the unexpected,'' in {\em The
  Basal Ganglia VIII} (J.~Bolam, C.~Ingham, and P.~Magill, eds.), ch.~1,
  pp.~1--24, Springer Science and Business Media, 2005.

\bibitem{Wilson1972}
H.~Wilson and J.~Cowan, ``Exictatory and inhibitory interactions in localized
  populations of model neurons,'' {\em Biophysical Journal}, vol.~12,
  pp.~1--24, 1972.

\bibitem{Lorenz1963}
E.~N. Lorenz, ``Deterministic nonperiodic flow,'' {\em Journal of the
  Atmospheric Sciences}, vol.~20, pp.~130--141, 1963.

\bibitem{Strogatz1994}
S.~H. Strogatz, {\em Nonlinear Dynamics and Chaos}.
\newblock Westview Press, 1994.

\bibitem{Hodgkin1952}
A.~L. Hodgkin and A.~F. Huxley, ``A quantitative description of membrane
  current and its application to conduction and excitation in nerve,'' {\em The
  Journal of Physiology}, vol.~117, no.~4, p.~500, 1952.

\bibitem{Mirollo1990}
R.~E. Mirollo and S.~H. Strogatz, ``Synchronization of pulse-coupled biological
  oscillators,'' {\em SIAM Journal on Applied Mathematics}, vol.~50, no.~6,
  pp.~1645--1662, 1990.

\bibitem{Ermentrout2002}
B.~Ermentrout, {\em Simulating, Analyzing, and Animating Dynamical Systems: A
  Guide to XPPAUT for Researchers and Students}.
\newblock Philadelpha, PA, USA: SIAM, 2002.

\bibitem{Brette2012}
R.~Brette and D.~F.~M. Goodman, ``Simulating spiking neural networks on gpu,''
  {\em Network: Computation in Neural Systems}, vol.~23, no.~4, pp.~167--182,
  2012.

\bibitem{Harris2004}
M.~J. Harris, {\em GPU Gems}, ch.~Fast Fluid Dynamics Simulation on the GPU,
  pp.~637--665.
\newblock Addison-Wesley, 2004.

\bibitem{Ahnert2011}
K.~Ahnert and M.~Mulansky, ``Odeint - solving ordinary differential equations
  in c++.'' http://arxiv.org/abs/1110.3397v1, 2011.

\bibitem{VanDerWalt2011}
S.~van~der Walt, S.~Colbert, and G.~Varoquaux, ``The numpy array: A structure
  for efficient numerical computation,'' {\em Computing in Science
  Engineering}, vol.~13, pp.~22--30, March 2011.

\bibitem{Klockner2012}
A.~{Kl{\"o}ckner}, N.~{Pinto}, Y.~{Lee}, B.~{Catanzaro}, P.~{Ivanov}, and
  A.~{Fasih}, ``{PyCUDA and PyOpenCL: A Scripting-Based Approach to GPU
  Run-Time Code Generation},'' {\em Parallel Computing}, vol.~38, no.~3,
  pp.~157--174, 2012.

\end{thebibliography}

\begin{appendices}
\renewcommand{\thefigure}{S\arabic{figure}}
\setcounter{figure}{0}

\section{Implementation Details}
In order to simulate a system of ODEs in Fireflies, the user must specify the $N$ \emph{State Variables}, along with the \emph{Parameters} of the system, and one or more \emph{Render Techniques} and \emph{Particle Groups}:

\begin{itemize}
\item{Each of the $N$ \emph{State Variables} corresponds to one of the system's ODEs and comprises a mathematical expression describing the variable's time derivative (right-hand side), along with the bounds of the variable. If a trajectory passes outside of bounds of any state variable during simulation it is reset to new random initial conditions.}
\item{For each \emph{Parameter} the user specifies a default value and the range of allowable values for the parameter.}
\item{A \emph{Render Technique} species a method for transforming the current position of each trajectory in phase space into a position of a particle on the screen. At present, particles can either be drawn using a simple 2D projection or a perspective 3D projection, and when specifying a projection the user must select which state variable corresponds to each of the (two or three) projection dimensions. The colour of rendered particles is also specified as part of a render technique; this can either be manually specified or varied automatically based on the position of the particle in 2D or 3D space.}
\item{Each \emph{Particle Group} consists of $P$ particles that share the same render technique and range of initial conditions. The user specifies: the number of particles; the render technique that should be used to draw them; whether the particles in the group should be simulated forwards or backwards in time; and, for each state variable, a range of valid initial conditions. These initial condition ranges define a cube in phase space from which a position for each particle in the group is chosen when the particle is first initialized (or reset as a result of going out of bounds).}
\end{itemize}

\begin{figure}[H]
\centering
\includegraphics[width=0.95\textwidth]{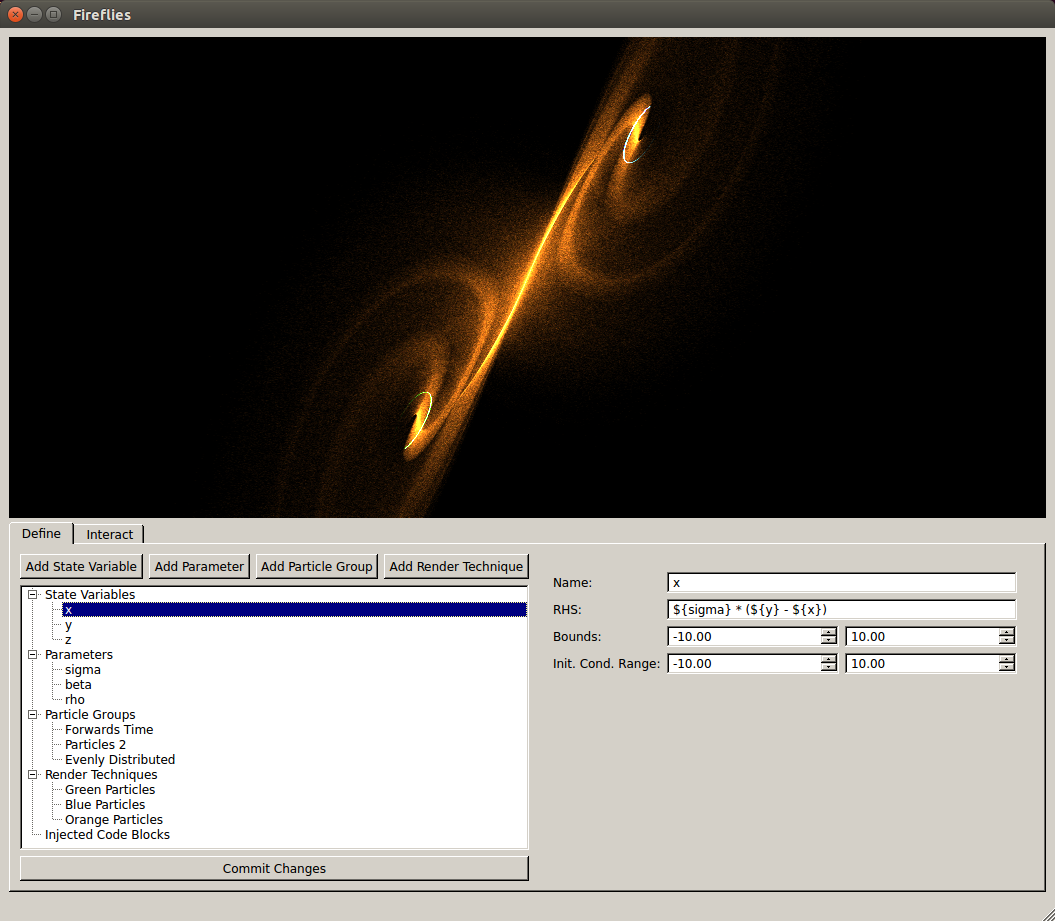}
\caption{A screenshot of the full Fireflies window showing the panel where the system to be simulated (the Lorenz equations in this case) is defined. Here, a state variable has been selected from the list of objects on the left hand side, allowing the properties of that state variable to be set using the panel on the right hand side.}\label{fig:sysdef_screenshot}
\end{figure}

\begin{figure}[H]
\centering
\includegraphics[width=0.95\textwidth]{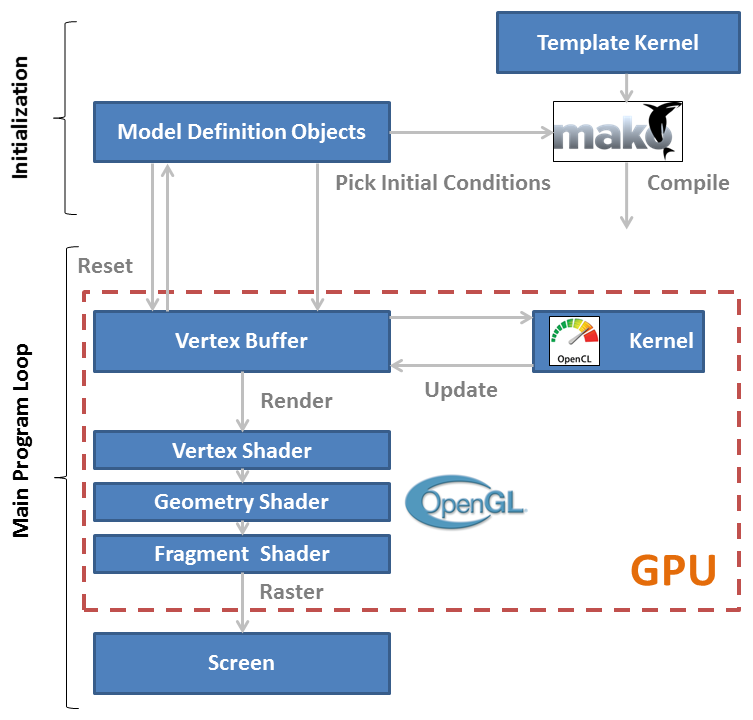}
\caption{A schematic overview of Fireflies. The processes at the top of the figure occur during intialization, and involve compiling an OpenCL kernel to advance the simulation, and generating random initial conditions. The main (lower) part of the figure shows the program's main loop, during which particles are repeatedly moved and then rendered to the screen. Note that almost all of the processing in the main loop occurs entirely within the GPU, avoiding the overhead of transferring data between the GPU and CPU. The exception to this is the resetting procedure, which only processes a subset of particles each frame.}\label{fig:software_overview}
\end{figure}

These objects that describe a system of equations are specified using the software's graphical user interface (Figure \ref{fig:sysdef_screenshot}). Once this has been done, Fireflies initializes the system before entering its main execution loop, which consists of repeatedly updating the particles' positions and rendering them to the screen. An overview of how the software is organised is shown in Figure \ref{fig:software_overview}. The following sections describe each stage of processing in more detail.

\subsection{Initialization}
When a system is first created, or subsequently modified, the GPU must be configured to simulate it. This involves populating the GPU's memory with the particles' initial positions and compiling and uploading programs to update and render the system onto the GPU. Setting up the particles' initial positions is straightforward: if the system is $N$-dimensional and consists of a total of $P$ particles, then an array with space for $N \cdot P$ single-precision floating point values is created and filled with random initial conditions (subject to the initial condition ranges specified by each particle's \emph{Group}) - this array is then transferred into the GPU's memory. Note that there are two possible ways of organising this memory: the first involves storing the $N$ values associated with the first particle, followed by those for the second particle, and so on (we call this ``row-major'' order), while the second method involves storing the $P$ values corresponding to the particles' positions in the first dimension, followed by the values for the second dimension, etc (we call this ``column-major'' order). At present Fireflies always uses row-major order, however it is likely that due to the way in which GPU memory access operates, column-major may gives significant performance benefits; this is something we plan to investigate in a future version of the software.

In order to simulate the system on the GPU, a program (or ``kernel'') must be compiled that updates the position of a single particle. The general form of this kernel is independent of the particular system of equations being integrated, and involves reading the particle's current position from memory, determining its new position using a 4th order Runge-Kutta routine, and then writing the new position back into memory. The only part of the kernel that changes for different systems of equations is that which calculates the time derivative for a given point in state space (the right hand side of each ODE). This function is generated automatically using the definitions for the state variables given by the user. In order to convert the specified system into kernel source code (in the ``OpenCL C'' language), Fireflies uses a template source code file which is processed by the Mako templating library\footnote{http://www.makotemplates.org} to produce the final kernel source. This source code is then compiled and uploaded onto the GPU using OpenCL.

In addition to the kernel, three other programs (or ``shaders'') must be compiled onto the GPU which perform the transformations necessary to render the particles to the screen. These shaders form a pipeline that converts from particle positions in state space into a series of squares (pairs of triangles) in two-dimensional screen space that can drawn by the graphics hardware. The different stages of this pipeline are the vertex shader, the geometry shader, and the fragment shader.

\begin{itemize}
\item{The \emph{vertex shader} takes as input the two or three components (depending on whether a 2D or 3D projection is required) of the particle's current position that correspond to the X, Y and Z directions of the scene to be rendered and outputs a two or three dimensional vector containing the position of the point transformed to be relative to an imaginary camera. This transformation is a simple linear transformation obtained by multiplying the vertex position by the ``Model-View'' matrix. In the case of a two-dimensional view, the camera always points directly at the plane containing the particles but it can be translated around and zoomed in and out. With a three-dimensional view, the user can interactively adjust the camera's position and orientation in space in order to ``fly around'' the system. As the user moves around the system, the Model-View matrix is updated accordingly.}

\item{The \emph{geometry shader} takes as input the camera-relative particle position from the vertex shader and first transforms this into two-dimensional position, using the ``View-Projection'' matrix. If the particle is to be rendered as a 2D projection then this transformation is straightforward, while for a 3D projection a standard perspective transformation matrix is used (for details of this see any 3D graphics programming resource). After the particle position has been projected into 2D, the geometry shader outputs four 2D vertices that correspond to a square that is centred on the particle position and facing the camera.}

\item{The \emph{fragment shader} is the final stage of the shader pipeline. When the graphics hardware renders the squares generated by the geometry shader on to the screen, it determines the colour of each pixel within the square by calling the fragment shader. The fragment shader generates an intensity value based on the distance from the centre of the square, producing a smooth circular particle shape. The actual colour output by the fragment shader is either specified explcitly by the particle's render technique, or varies linearly as a function of the particle's position in state space. The colour of the pixel on the screen is set to its current colour plus the colour contributed by the particle being drawn\footnote{This corresponds to \textit{glBlendFunc(GL\_SRC\_ALPHA, GL\_ONE)}.}.}
\end{itemize}

\subsection{Running the Simulation}
Once the system has been specified and the GPU initialized with the kernel, shaders and initial particle positions, the simulation can be started by making the step size non-zero (both positive and negative step sizes are possible, to simulate the system either forwards or backwards in time). 

When the simulation is running, the main simulation loop repeatedly instructs the GPU to alternately execute the kernel and then the shader pipeline, causing the particles to be updated and drawn to the screen. The user can see, in real time, how the particles are moving around the state space and over time the system's attractors become clear. The user can explore the state space by changing the position of the camera in 2D or 3D space (effectively changing the Model-View matrix in GPU memory) using the mouse and keyboard. The system's parameters can also be changed during simulation by dragging sliders on the user interface, which updates the location in GPU memory where the corresponding parameter value is stored.

Some operations, such as checking for and resetting particles that have moved outside the bounds of the system to new random positions are not well-suited for GPU processing; Fireflies performs these operations on the CPU. Since copying data between the CPU and GPU is relatively slow, it is not practical to copy the positions of all of the particles back and forth at every time step. Instead, only a fraction of the particle position data is copied and processed by the CPU each frame. The CPU processes this batch of particles by checking if any of them have left the bounds of the system and chooses new initial conditions where necessary, before copying the affected particles' positions back to the GPU. This method ensures that simulations run at a smooth frame rate, although it means that particles are typically reset several time steps after their reset conditions are first satisfied. The user can adjust the size of CPU processing batches while the simulation is running, in order to find the setting that maximizes performance.

\section{Supplementary Movies}
\begin{itemize}
\item \texttt{stngpe\_2d.mp4}: Simulation of the two-dimensional STN-GPe model, showing the bifurcations that occur as the parameter $w_{ss}$ is increased (refer to main text for details).
\item \texttt{stngpe\_bif.mp4}: Interactive three-dimensional bifurcation diagram for the STN-GPe model, with the parameter $w_{ss}$ plotted as the third dimension.
\item \texttt{lorenz.mp4}: The Lorenz system visualized in three dimensions. Initially the parameter $r$ has a low value and the system has two stable spirals. After some time, the parameter is changed to its ``classical'' value of 28, causing the spirals to become unstable and a strange attractor to appear. Particle colours as in Figure \ref{fig:lorenz_composite}.
\item \texttt{hh\_combined.mp4}: Three Hodgkin-Huxley type neurons coupled by excitatory synapses. The position of each particle in three dimensional space corresponds to the membrane potential of each neuron. After an initial period of unsynchronized firing, the particles all eventually approach of several limit cycles corresponding to different spiking patterns (see main text).
\end{itemize}
\end{appendices}
\end{document}